\documentclass[12pt]{article}

\def\nn{\nonumber}

\def\nl{\hfil\break} 

\def\nt{\noindent} 
\def\({\left(}
\def\){\right)}
 
\def\id{{\bf 1}} 
\def\va{{\vert a\vert}}

\def\td{{\tilde d}} 
\def\had{{\textstyle{d\over2}}} 
\def\half{{\textstyle{1\over2}}} 
\def\tdd{{[\had]}}

\def\thd{{\hat d}}
\def\hadd{{\textstyle{d+1\over2}}} 
\def\ttd{{[\hadd]}}

\def\tp{{\tilde p}} 
\def\hap{{\textstyle{p\over2}}} 
\def\tpp{{[\hap]}}
\def\tD{{\hat D}} 

\def\rank{{\rm rank\,}}
 
\def\D{\Delta} 

\def\bac{{C\kern-5.2pt I}} \def\bbc{{C\kern-6.5pt I}} 
\def\bbr{I\!\!R}  
\def\a{\alpha} \def\b{\beta} \def\g{\gamma} \def\d{\delta}

 \def\l{\lambda} 
  \def\cd{{\cal D}} 
\def\tcd{\tilde{\cal D}} 
\def\cc{{\cal C}}  \def\tmu{{\tilde \mu}} 
\def\hc{\hat{C}}  
 
\def\tih{{C_\chi}} \def\tch{{\tilde{\chi}}} \def\tcp{{\tilde{\chi'}}}

\def\htt{\hat{T}} \def\tt{{{\tau}}} \def\ttp{{{\tau'}}}

\def\cm{{\cal M}}

  \def\cs{{\cal S}} 
\def\co{{\cal O}} 
\def\cg{{\cal G}}  
 \def\hd{{D}}

 \def\y{\eta} 
  
\def\lra{\leftrightarrow} 
\def\rra{\longrightarrow} \def\lla{\longleftarrow}

\def\tn{{\tilde n}} \def\tN{{\tilde N}} 
\def\t{\tau} 
\def\kc{K_\chi^\t}

\newcommand{\ben}{\begin{enumerate}}
\newcommand{\een}{\end{enumerate}}
\newcommand{\be}{\begin{equation}}
\newcommand{\ee}{\end{equation}}
\newcommand{\bea}{\begin{eqnarray}}
\newcommand{\eea}{\end{eqnarray}}
\newcommand{\bc}{\begin{center}}
\newcommand{\ec}{\end{center}}

\def\eqn{\be\label}
\def\eqnn{\bea\label}

\begin{document}

\rightline{IC/2002/33;\ \ {hep-th/0207116;\ \ {}May 2002}}

\begin{center}
\vspace*{1.0cm}

{\LARGE{\bf AdS/CFT Correspondence and Supersymmetry}} 
\footnote{Lectures at the 1st Summer School in Modern 
Mathematical Physics, Sokobanja, Yugoslavia, 13-25.8.2001;  
to appear in the Proceedings of the School.}

\vskip 1.5cm

{\large {\bf V.K. Dobrev}} 

\vskip 0.5 cm 

Institute of Nuclear Research and Nuclear Energy\\ 
Bulgarian Academy of Sciences\\ 72 Tsarigradsko Chaussee, 1784 
Sofia, Bulgaria\footnote{Permanent address.} 

and 

The Abdus Salam International Center for Theoretical Physics\\  
Strada Costiera 11, P.O. Box 586,\\  
34100 Trieste, Italy

\end{center}

\vspace{1 cm}

\begin{abstract}
We use the group-theoretic interpretation of the AdS/CFT
correspondence which we proposed earlier in order to lift
intertwining operators acting between boundary conformal
representations to intertwining operators acting between bulk
conformal representations. Further, we present the classification
of the positive energy (lowest weight) unitary irreducible
representations of the $D=6$ superconformal algebras
$osp(8^*/2N)$. 
\end{abstract}

\section{Introduction} 
Recently there was renewed interest in (super)conformal field
theories in arbitrary dimensions. This happened after the
remarkable proposal in \cite{Malda}, according to which the large
$N$ limit of a conformally invariant theory in $d$ dimensions is
governed by supergravity (and string theory) on $d+1$-dimensional
$AdS$ space (often called $AdS_{d+1}$) times a compact manifold.
Actually the possible relation of field theory on $AdS_{d+1}$ to
field theory on $\cm_d$ has been a subject of long interest, cf.,
e.g., \cite{FlFr,NiSe,GNW}, and also \cite{FeFr} for discussions
motivated by recent developments, and additional references. The
proposal of \cite{Malda} was elaborated in \cite{GKP} and
\cite{Wi} where was proposed a precise correspondence between
conformal field theory observables and those of supergravity:~
correlation functions in conformal field theory are given by the
dependence of the supergravity action on the asymptotic behavior
at infinity. More explicitly, a conformal field ~$\co$~
corresponds to an AdS field ~$\phi$~ when there exists a
conformal invariant coupling ~$\int \phi_0\, \co$~ where
~$\phi_0$~ is the value of ~$\phi$~ at the boundary of
$AdS_{d+1}\,$ (\cite{Wi}). Furthermore, the dimension ~$\D$~ of
the operator ~$\co$~ is given by the mass of the particle
described by ~$\phi$~ in supergravity \cite{Wi}. After these
initial papers there was an explosion of related research of
which of interest to us are two aspects: 1) calculation of
conformal correlators from AdS (super)gravity and various
questions of holography, cf., e.g., [8-78];
\ 2) matching of supergravity and superstring spectra with
superconformal theories, cf., e.g., [79-140]. 

For the ~{\it first} aspect of the AdS/CFT correspondence 
one of the main features furnishing it is that 
the boundary $\cm_d$ of $AdS_{d+1}$ is in fact
a copy of $d$-dimensional Minkowski space (with a cone added 
at infinity); the symmetry group $SO(d,2)$ of $AdS_{d+1}$ acts
on $\cm_d$ as the conformal group. The fact that 
$SO(d,2)$ acts on $AdS_{d+1}$ as a group of ordinary symmetries 
and on $\cm_d$ as a group
of conformal symmetries means that there are two ways to
get a physical theory with $SO(d,2)$ symmetry:
in a relativistic field theory (with or without gravity) on 
$AdS_{d+1}$, or in a conformal field theory on $\cm_d$. 

In an earlier paper \cite{Doads} we gave a group-theoretic interpretation 
of the above correspondence. [In fact such an interpretation is partially 
present in \cite{DMPPT} for the $d=3$ Euclidean version of the 
AdS/CFT correspondence in the context of the construction of discrete 
series representations of the group $SO(4,1)$ involving 
symmetric traceless tensors of arbitrary nonzero spin.] 
In short the essence of our interpretation is that the above 
correspondence is a relation of ~{\it representation 
equivalence}~ between the representations describing the fields 
~$\phi$, ~$\phi_0$~ and ~$\co$. There are actually two kinds of 
equivalences. The first kind is new (besides the example from 
\cite{DMPPT} mentioned above) and was proved in \cite{Doads} - it is 
between the representations describing the bulk fields and the 
boundary fields. The second kind is well known - it is the equivalence 
between boundary conformal representations which are 
related by restricted Weyl reflections, the representations here being the 
coupled fields ~$\phi_0$~ and ~$\co$. 
 Our interpretation means that the operators relating these fields 
are ~{\it intertwining operators}~ between 
(partially) equivalent representations. Operators giving the first 
kind of equivalence for special cases were actually given in, e.g., 
\cite{Wi,HeSf,MuVi,LiTs,CVY} - in \cite{Doads} they were constructed 
in a more general setting 
from the requirement that they are intertwining 
operators. The operators giving the second kind of equivalence are 
provided by the standard conformal two-point functions. 
Using both equivalences we have found that the bulk field has naturally 
~{\it two}~ boundary fields, namely, the coupled fields 
~$\phi_0$~ and ~${\cal O}$, the limits being governed by the 
corresponding conjugated conformal weights ~$d-\D$ and ~$\D$. 
Thus, from the point of view of the 
bulk-to-boundary correspondence 
the coupled fields ~$\phi_0$~ and ~${\cal O}$ are 
generically\ {\it on an equal footing.} [The appearance of two 
boundary fields was used later in \cite{KlWi} in a slightly 
different context, namely, that the theory with the same 
classical AdS Lagrangian can be interpreted in terms of 
two different CFT's with the conjugated dimensions. 
This is possible only for sufficiently negative
AdS-mass-squared, so that both dimensions would not 
be lower than the unitarity bound.] 

In the present paper we review also the results of \cite{Dowig6} 
in order to lift intertwining operators acting 
between boundary conformal representations to 
intertwining operators acting between bulk 
conformal representations. 

For the ~{\it second} aspect of the AdS/CFT correspondence one
of the most important tasks is the
classification of the UIRs of the corresponding 
superalgebras. 
Particularly important are those for ~$D\leq 6$~ since in these
cases the relevant superconformal algebras satisfy \cite{Nahm} the
Haag-Lopuszanski-Sohnius theorem \cite{HLS}. 
Until recently such classification was known only for the ~$D=4$~
superconformal algebras ~$su(2,2/N)$ \cite{FF} (for $N=1$),
\cite{DPm,DPu,DPf,DPp}. 
Recently, the classification for ~$D=3,5,6$~ was given \cite{Min}
but the results were conjectural and there was not enough detail in
order to check these conjectures. In view of the interesting
applications \cite{FSa,EFS,FSb} of $D=6$ unitary
irreps to the analysis of OPEs and 1/2 BPS operators we decided
to reexamine the list of UIRs of the ~$D=6$~ superconformal
algebras ~$osp(8^*/2N)$ in detail \cite{Dosu6}. We confirm all
but one of the conjectures of \cite{Min} and thus, we give the final
list of UIRs for ~$D=6$. Our main tool is the explicit
construction of the norms. This, on the one hand, 
enables us to prove the unitarity list, and, on the other hand, 
enables us to give explicitly the states of the irreps. 
We give a brief summary of \cite{Dosu6} in the last section here.

\section{Conformal field theory representations} 
As in \cite{Doads} we consider the 
Euclidean version of the AdS/CFT correspondence. 
For definiteness we use the following defining relation of 
the Sitter group ~$G$~: 
\eqnn{dsg} G ~=~ \{ g \in GL(d+2, \bbr) ~\vert ~ 
^t g\y g = \y \doteq {\rm diag}(-1,\dots,-1,1), \nn\\
~~\det g =1, ~~g_{d+2,d+2} \geq 1 \} \eea 
Thus, ~$G=SO_e(d+1,1)$, i.e., it 
is the identity component of ~$O(d+1,1)$, 
($^t g$ is the transposed of $g$). Note that for ~$d~~even$~ 
some expressions are simpler if we work with the 
extended de Sitter group:
\eqn{ddsg} G' ~\doteq~ \{ g \in GL(d+2, \bbr) ~\vert ~ 
^t g\y g = \y, ~~g_{d+2,d+2} \geq 1 \} \ee 
which includes reflections of the first $d+1$ axes. 
The representations of $G$ used in conformal field theory 
are called (in the representation theory of semisimple Lie groups) 
generalized principal series representations (cf. \cite{Kn}). 
In \cite{DMPPT,DP,Dob} 
they were called ~{\it elementary representations} (ERs). 
They are obtained by induction from the subgroup$\,$ $P=MAN$, where 
~$M=SO(d)$~ is the Euclidean Lorentz group, 
~$A$~ is the one-dimensional dilatation group, 
~ $N$~ is the group of special conformal transformations 
(isomorphic to ~$\bbr^d$), ~$P$ is called a parabolic subgroup of $G$. 
The induction is from 
unitary irreps of $M=SO(d)$, from arbitrary (non-unitary) characters of $A$, 
and trivially from $N$. There are several realizations of these 
representations. We give now the so-called ~{\it noncompact picture}~ 
of the ERs - it is the one actually used in physics. 

The representation space of these induced representations 
consists of smooth functions on ~$\bbr^d$~ with values in 
the corresponding finite-dimensional representation space of ~$M$, i.e.: 
\eqn{funn} C_\chi ~=~ \{ f \in C^\infty(\bbr^d,V_\mu)\} \ee 
where$\,$ $\chi\, =\, [\mu,\D]$,$\,$ ~$\D$~ is the conformal weight, 
~$\mu\,$ is a unitary irrep of $M$,$\,$ 
$V_\mu\,$ is the finite-dimensional representation space of $\mu$. 
In addition, these functions have special 
asymptotic expansion as ~$x\to \infty$. The leading term 
of this expansion is ~$f(x) ~\sim~ {1\over (x^2)^\D}\, f_0$, 
(for more details we refer to \cite{DMPPT,DP,Dob}). 
The representation ~$T^\chi$~ acts in $\tih$ by: 
\eqn{lart} (T^\chi(g) f) (x) ~=~ \va^{-\D} \cdot \hd^\mu(m)\, 
f ( x') \ee 
where the nonglobal Bruhat decomposition $g=\tn m a n$ is 
used:
\eqn{nnam} g^{-1}\tn_x ~=~ \tn_{x'} m^{-1}{a}^{-1} n^{-1} ~, \quad 
g\in G , \, \tn_x, \tn_{x'} \in \tN , \, m\in M,\, a\in A,\, n\in N \ee
where ~$\tN$~ is the abelian group of 
Euclidean translations (isomorphic to ~$\bbr^d$), 
$\hd^\mu(m)$ is the representation matrix of $\mu$ in $V_\mu\,$.

Note that the representation data given by ~$\chi=[\mu,\D]$~ fixes also 
the value of the Casimir operators ~$\cc_i$~ 
in the ER$\,$ $C_\chi\,$, independently of the latter reducibility. 
For later use we write:
\eqn{cas} \cc_i\, f(x) ~=~ \l_i(\mu,\D)\, f(x) ~, \qquad 
i=1,\dots,\rank G ~=~ [\had]+1, \ee

Next, we would like to recall the general expression of 
the conformal two-point function ~$G_\chi(x_1-x_2)$~ 
(for special cases cf. \cite{Pol,FGGP,Mig,FrPa}, for the 
general formula with special stress on the role of the 
conformal inversion, cf. \cite{Kol}, also \cite{DMPPT}):
\eqnn{tpf} 
&& G_\chi(x) ~=~ {\g_\chi\over (x^2)^\D }\, \hd^\mu(r(x)) \\ [2mm]
&& r(x) ~=~ \pmatrix{\tilde r(x) 
& 0 & 0 \cr 
0 & 1 & 0 \cr 0 & 0 & 1 } \in M ~, \quad 
\tilde r(x) ~=~ \left( {2\over x^2}\, x_i x_j - \d_{i j} \right) \nn\eea 
where ~$\g_\chi$~ is an arbitrary constant for the moment. 
(Note that for ~$d$~~ {\it even}~ $r(x)$ $\in O(d)$, so we work
with ~$G'$, cf. (\ref{ddsg}).) 

Finally, we note the 
intertwining property of ~$G_\chi(x)$. Namely, let ~$\tch$~ be 
the representation conjugated to ~$\chi$~ by a restricted 
Weyl reflection, i.e., by the nontrivial element of the 
restricted Weyl group $W(G,A)$ \cite{DMPPT}. Then we have: 
\eqn{chcj} \tch ~\doteq~ [\, \tmu,\, d-\D\, ] ~, \quad {\rm for} ~
\chi = [\mu,\D], \ee 
where ~$\tmu$~ is the ~{\it mirror image}~ representation of 
~$\mu$. (For ~$d$~ odd ~$\tmu \cong \mu$, while for ~$d$~ even ~$\tmu$~ 
may be obtained from ~$\mu$~ by exchanging the representation 
labels of the two distinguished Dynkin nodes of ~$SO(d)$.) 
Then there is the following intertwining operator \cite{Kol,DMPPT}:
\eqnn{inti}
&&G_\chi ~:~ C_\tch \rra C_\chi ~, \qquad 
T^\chi (g) \circ G_\chi ~=~ G_\chi \circ T^\tch (g) ~, \quad \forall g ~, 
 \\ [2mm]
&&(G_\chi f) (x_1) ~\doteq ~ \int G_\chi(x_1-x_2)\, f(x_2)\, d x_2 ~,
\quad d x \equiv d^d x ~,\nn \\ [2mm] 
&&C_\chi ~\simeq ~C_\tch \eea 
the last line uses our symbolic notation for partial equivalence 
between $\chi$ and $\tch$. Note that because of this equivalence the 
values of all Casimirs coincide:
\eqn{csmr} \l_i(\tmu, d-\D) ~=~ \l_i(\mu, \D) ~, \qquad \forall i ~. \ee

\section{Representations on the bulk space} 
In the previous section we discussed representations 
on ~$\bbr^d\cong \tN$~ induced from the parabolic subgroup ~$MAN$~ 
which is natural since the abelian subgroup ~$\tN$~ is locally 
isomorphic to the factor space ~$G/MAN$ (via the Bruhat decomposition). 
Similarly, it is natural to discuss 
representations on the bulk space ~$\cs\cong \tN A$~ 
which are induced from the maximal compact subgroup ~$K=SO(d+1)$~ 
since the solvable group ~$\tN A$~ is isomorphic to the factor space ~$G/K$ 
via the global ~{\it Iwasawa}~ decomposition ~$G ~=~ \tN AK$ 
(cf. the details in \cite{Doads}\footnote{Note that in \cite{Doads}
the bulk space ~$\cs\cong \tN A\cong SO(d+1,1)/SO(d+1) $~ was 
called de Sitter space, though in the literature the latter name
is used for the space ~$SO(d+1,1)/SO(d,1)$. The latter space was 
used recently for extensive study of the so-called ~{\it dS/CFT}~
correspondence, cf., e.g., \cite{Strominger,Klemm,PeSi,Myung}.}). 
Namely, we consider the representation spaces:
\eqn{funk} \hc_\t ~=~ \{ \phi \in 
C^\infty(\bbr^d\times\bbr_{>0}\,,U_\t) \} \ee 
where$\,$ $\t\,$ is an arbitrary unitary irrep of $K$,$\,$ 
$U_\t\,$ is the finite-dimensional representation space of $\t$, 
with representation action:
\eqn{lartu} 
(\htt^\t(g)\phi) (x,\vert a\vert) ~=~ \tD^\t (k)\, \phi ( {x'}, 
\vert a'\vert) \ee 
where the Iwasawa decomposition is used: 
\eqn{nak} g^{-1}\tn_x a ~=~ \tn_{x'} a' k^{-1} ~, \quad 
g\in G ,\, k\in K, \, \tn_x, \tn_{x'} \in \tN , \, a, a' \in A \ee 
and $\tD^\t(k)$ is the representation matrix of $\t$ in $U_\t\,$. 
However, unlike the ERs, these representations are reducible, 
and to single out an irrep equivalent, say, a 
subrepresentation of an ER, one has to look for solutions 
of the eigenvalue problem related to the Casimir operators. 

In the actual implementation of (\ref{lartu}) and (\ref{nak}) we 
use the following paramet\-rization of ~$k$: 
\eqn{mtra} k ~=~ \pmatrix{ k_{i j} & k_{i,d+1} & 0 \cr 
k_{d+1,j} & k_{d+1,d+1} & 0 \cr 
0&0&1 \cr} \in K ~, \quad \left( k_{\a\b} \right) \in SO(d+1) ~.\ee 
Further we shall need also the unique decomposition:
\eqn{dkk}
k ~=~ m(k) k_f ~, \qquad 
m(k)=\pmatrix{\tilde m(k)& 0 &0 \cr 0&1&0 \cr 0&0&1} \in M ~, \qquad 
k_f=\pmatrix{\tilde k_f& 0 \cr 0&1} \in K \ee 
representing the decomposition of 
~$K$~ into its subgroup ~$M$~ and the coset ~$K/M$~: ~
~$K ~\cong ~ M~K/M$. Explicitly, 
we have (for $k_{d+1,d+1}\neq -1$):
\eqnn{dek} 
&& \tilde m(k) ~=~ \left( k_{i j} - 
{1\over 1+k_{d+1,d+1}}\, k_{i,d+1}\, k_{d+1,j} 
\right) \nn\\ [2mm]
&& \tilde k_f ~=~ 
\pmatrix{ \d_{i j} - {2\over 1+x^2}\, x_{i} x_{j} 
& - {2\over 1+x^2}\, x_{i}\cr 
{2\over 1+x^2}\, x_{j} & {1-x^2\over 1+x^2}} ~\doteq ~\tilde k_x 
~, ~~~ x\in ~\bbr^d ~, \eea 
$x_i ~=~ k_{d+1,i}/ ( 1+k_{d+1,d+1} )$. 
Note that ~$k_x ~\doteq ~\pmatrix{\tilde k_x& 0 \cr 0&1}$~ 
appeared in (1.30a) of \cite{DMPPT}. 

Further, we would like to extract from ~$\hc_\t$~ 
a representation that may be equivalent to ~$C_\chi\,$, $\chi=[\mu,\D]$. 
The first condition for this is that the ~$M$-representation ~$\mu$~ is 
contained in the restriction of the ~$K$-representation ~$\t$~ to ~$M$. 
Another condition is that the two representations 
would have the same Casimir values ~$\l_i(\mu,\D)$. Having in mind 
the degeneracy of Casimir values for partially equivalent 
representations (e.g., (\ref{csmr})) we add also the appropriate 
asymptotic condition. Furthermore, from now on we shall suppose 
that ~$\D$~ is real. Thus, we shall use the representations: 
\eqnn{funr} 
&& \hc^\t_\chi ~=~ \{\, \phi \in \hc_\t ~:~~ 
\cc_i\, \phi (x,\va) ~=~ \l_i(\mu,\D)\, \phi (x,\va) ~,
\quad \forall i ~, \quad 
\mu\in\t\vert_M ~, \nn\\ [2mm] 
&&\phi (x,\va) ~\sim~ \va^\D\, \varphi(x) ~ {\rm for}~ \va\to 0\, 
\} \eea 
Unlike their action on the ERs $C_\chi\,$ the Casimirs ~$\cc_i$~ 
here are differential operators and the elements of 
~$\hc^\t_\chi$~ are solutions of the equations in (\ref{funr}). 

In \cite{Doads} it was shown that generically 
the functions in (\ref{funr}) have also 
a second limit with ~$\D\to d-\D$~: 
\eqn{bndd} \tilde \varphi(x) ~=~ \lim_{\va\to 0}\, \va^{\D-d}\, 
\phi (x, \va) \ee
For generic representations this establishes 
the following important relation:
\eqn{dud} \hc^\t_\chi ~=~ \hc^\t_\tch ~, \qquad \chi =[\mu,\D], ~
\tch =[\tmu,d-\D] ~. \ee 
For non-generic representations the second asymptotic expansion 
of $\phi$ contains logarithms (cf., e.g., (7.45) of \cite{DMPPT} 
and \cite{FMMR,CKA}), 
and then the representations $\chi$ and $\tch$ are only partially 
equivalent:
\eqn{pdud} \hc^\t_\chi ~ \simeq ~ \hc^\t_\tch ~. \ee

 \section{Intertwining relations between conformal and 
bulk representations}
This Section contains the main result of \cite{Doads}, 
explicating the relations between 
CFT and bulk representations as intertwining relations. 
We first give the intertwining operator 
from the bulk to the CFT realization. 
The operator is mapping a 
function on the bulk space to its boundary value and was 
used in a restricted sense (explained below) 
in many papers, starting from~\cite{Wi}. 

\nt 
{\bf Theorem:}~~ {\it Let us define the operator: 
\eqn{aaa} L_\chi^\t ~:~ \hc^\t_\chi ~\rra ~C_\chi \, , \ee
with the following action:
\eqn{aab} (L_\chi^\t \phi ) (x) ~=~ \lim_{\va\to 0}\ 
\va^{-\D}\ \Pi^\t_\mu\ \phi (x, \va) \ee 
where ~$\Pi^\t_\mu$~ is the standard projection operator from the 
~$K$-representation space ~$U_\t$~ to the ~$M$-representation space
~$V_\mu\,$, which acts in the following way on the 
$K$-representation matrices: 
\eqn{prj} \Pi^\t_\mu\ \tD^\t (k) ~=~ \hd^\mu (m(k))\ \Pi^\t_\mu \ 
\tD^\t (k_f) \ee 
where we have used (\ref{dkk}). 
Then ~$L_\chi^\t$~ is an intertwining operator, i.e.: 
\eqn{aac} L_\chi^\t \circ \htt^\t (g) ~=~ 
 T^\chi (g) \circ L_\chi^\t ~, \quad \forall g\in G ~. \ee 
In addition, in (\ref{aab}) the operator ~$\Pi^\t_\mu$~ 
acts in the following truncated way:} 
\eqn{prja} \Pi^\t_\mu\ \tD^\t (k) ~=~ \hd^\mu (m(k))\ \Pi^\t_\mu \ee
The proof is given in \cite{Doads}.

\medskip 

Next we consider the operator inverse to ~$L_\chi^\t\,$~ which 
would restore a function on the bulk space from its boundary 
value, as discussed in [7-18]. 
Again what was new in \cite{Doads} was that the operator was defined 
as intertwining operator between 
exactly defined spaces in a more general setting. 
Moreover, the operator was constructed just from 
the condition that it is an intertwining integral operator. 
Namely, we started with the operator: 
\eqn{inv} {\tilde L}_\chi^\t 
~:~ C_\chi ~\rra ~ \hc_\chi^\t ~,\ee 
using the following Ansatz:
\eqn{inta} \( {\tilde L}_\chi^\t \ f\) (x,\va) ~=~ 
\int \kc (x,\va;x')\, f(x')\, d x' \ee 
where ~$\kc (x,\va;x')$~ is a linear operator acting 
from the space $V_\mu$ to the space $U_\t\,$, 
and supposed that 
~${\tilde L}_\chi^\t$~ is an intertwining operator, i.e.: 
\eqn{iaac} \htt^\t (g) \circ {\tilde L}_\chi^\t 
~=~ {\tilde L}_\chi^\t \circ T^\chi (g) ~, \quad \forall g\in G ~. \ee
{}From this we obtained (cf. the details in \cite{Doads}) that $\kc$ is fixed 
up to an overall multiplicative constant $N_\chi^\t$ and explicitly is: 
\eqnn{fff} 
&&\kc (x,\va;x') ~=~ \kc (x-x',\va) \nn\\ [2mm]
&&\kc (x,\va) ~=~ N_\chi^\t\ \( {\va \over x^2 +\va^2}\)^{d-\D}\ 
\tD^\t (k_{-{x\over \va}})\ \Pi^\mu_\t \eea 
where ~$\Pi^\mu_\t$~ is the canonical embedding operator from $V_\mu$ to 
$U_\t\,$, such that ~$\Pi^\t_\mu \circ \Pi^\mu_\t ~=~ \id_\mu\,$, 
and $k_x$ is given in (\ref{dek}). 

The above operator exists for arbitrary representations 
~$\t$~ of ~$K=SO(d+1)$~ which contain 
the representation ~$\mu$~ of ~$M=SO(d)$. 
We use the standard ~$SO(p)$~ representation 
parametrization: ~$[\ell_1,\dots,\ell_\tp]$, ($\tp\equiv\tpp$), 
where all $\ell_j$ are simultaneously integer or 
half-integer, all are positive except for $p~~even$~ when 
~$\ell_1$~ can also be negative, and they are ordered: 
~$\vert \ell_1\vert\leq \ell_2\leq \dots \leq \ell_\tp$. 
The condition that ~$\t ~=~ [\ell'_1,\dots,\ell'_\thd]$, 
($\thd \equiv \ttd$), contains ~$\mu ~=~ [\ell_1,\dots,\ell_\td]$, 
($\td \equiv \tdd$), explicitly is:
\eqnn{cont}
&&\vert\ell'_1\vert \leq \ell_1 \leq \dots \leq \ell_\td 
\leq \ell'_\thd ~, \qquad d~~ odd, ~~\thd=\td+1 \nn\\ [2mm]
&&-\ell'_1\leq \ell_1 \leq \ell'_1\dots \leq \ell_\td \leq 
\ell'_\thd ~, \qquad d~~ even, ~~\thd=\td \eea 

If one is primarily concerned with the ERs ~$\chi = [\mu,\D]$~ 
it is convenient to chose a 'minimal' representation 
~$\t(\mu)$~ of ~$K=SO(d+1)$~ containing ~$\mu\,$. 
This depends on the parity of ~$d$. Thus, for ~$\mu$~ as above, 
when ~$d$~ is ~{\it odd}~ we would choose: 
\eqn{ppp} \t(\mu) ~=~ [\ell_1,\ell_1,\dots,\ell_\td] 
\qquad {\rm or}\qquad \tilde\t(\mu) ~=~ 
[-\ell_1,\ell_1,\dots,\ell_\td] ~, \qquad \mu \cong \tmu ~, \ee
while for ~{\it even}~ $d$~ we would choose: 
\eqn{pppp} \t(\mu) ~=~ [\vert\ell_1\vert,\ell_2\dots,\ell_\td] 
~=~ \t(\tilde\mu) ~\cong~ \tilde\t(\mu) ~=~ \tilde\t(\tilde\mu) 
~, \quad \tilde\mu ~=~ [-\ell_1,\ell_2\dots,\ell_\td] ~. \ee
Thus, in the odd $d$ case for each ~$\mu$~ we would choose between 
two $K$-irreps which are mirror images of one another, 
while in the even $d$ case to each two mirror-image irreps of $M$ 
we choose one and the same irrep of $K$.

The explicit examples which appeared in the literature are 
actually in the cases in which ~$\t=\t(\mu)$, e.g., 
\cite{Wi,HeSf,MuVi,LiTs,CVY}, though there is no such 
interpretation as we have here.

\section{Intertwining operators on the bulk space} 
In this section we review \cite{Dowig6}. We 
show how to lift intertwining operators acting 
between boundary conformal representations to 
intertwining operators acting between bulk 
conformal representations. Of course, for generic representations 
there is nothing to do, since the pairs of equivalent 
boundary representations $\chi$ and $\tch$ are equivalent to the 
same bulk representation. However, for nongeneric boundary 
representations when $\chi$ and $\tch$ are partially equivalent 
but not equivalent, the situation is much more interesting. 
In this case besides the pair $\chi_0\equiv \chi$ and 
$\tch_0\equiv\tch$ there 
exist ~$\td$~ more such pairs $\chi_i$ and $\tch_i$ ($i=1,...,\td$) 
so that these $2\td+2$ ERs are partially equivalent between themselves, 
(for the explicit parametrization of these ERs we refer to 
\cite{DMPPT,DP} for early partial cases, and \cite{Dobtw} 
for the general case). These partial equivalences are realized 
by ~$2\td$~ intertwining differential operators. The latter come 
in pairs, i.e., if there exists an intertwining differential operator 
~$\cd$~ acting from ~$C_{\chi_j}$~ to ~$C_{\chi_{j'}}$, ($j,j'=0,1,...,\td$;\ 
$j\neq j'$), then there exists an intertwining differential operator ~$\cd'$~ 
acting from ~$C_{\tch_{j'}}$~ to ~$C_{\tch_{j}}$. 

Now we shall use 
these operators and the operators bulk$\lra$boundary operators 
of the previous section to build operators acting between 
bulk representations. For notational simplicity we write $\chi,\chi'$ 
instead of $\chi_j\,,\chi_{j'}\,$. 
We start with the operator ~$\cd$~ 
acting from ~$C_\chi$~ to ~$C_{\chi'}\,$, where ~$\chi' = [\mu',\D']$, 
and lift it to an operator ~$\tcd$~ acting from ~$\hc_\chi^\t$~ to 
~$\hc_{\chi'}^\ttp$, where ~$\t'$~ is a UIR of $K$ containing $\mu'$. 
Explicitly, we have:
\eqn{inb} \tcd ~:~ \hc_\chi^\t ~\rra~ \hc_{\chi'}^\ttp ~, \qquad 
\tcd ~=~ \tilde L_{\chi'}^\ttp\ \circ \cd \circ L_\chi^\t \ee 
Analogously, we have for the operator ~$\tcd'$~ acting from 
~$\hc_\tcp^\ttp$~ to ~$\hc_\tch^\t$~:
\eqn{inbb} \tcd' ~:~ \hc_\tcp^\ttp ~\rra~ \hc_\tch^\t ~, \qquad 
\tcd' ~=~ \tilde L_{\tch}^\t\ \circ \cd' \circ L_\tcp^\ttp \ee 
The explicit parametrization and expressions 
of all possible operators $\tcd,\tcd'$ will be given elsewhere. 

Finally, we notice that all operators that we have used 
may be found on the following diagram:
\eqn{dgr} 
\matrix{&&&&\cr 
&&\hc_\tch^\t\ \simeq\ \hc_\chi^\t &&\cr 
&&&&\cr &&&&\cr 
&L_\tch^\tt\ \swarrow\nearrow 
\tilde L_\tch^\tt 
&& \tilde L_\chi^\t\ 
\nwarrow\searrow
L_\chi^\t & \cr
&&&&\cr &&&&\cr 
&& G_\tch && \cr
C_\tch &&{\lla\atop\rra} && C_\chi\cr 
&& G_\chi && \cr 
&&&&\cr &&&&\cr 
\cd'\, \uparrow\, \phantom{\cd'} &&&&\phantom{\cd}\, \downarrow\, \cd 
&&&&\cr &&&&\cr
&& G_\tcp && \cr
C_\tcp &&{\lla\atop\rra} && C_{\chi'}\cr 
&& G_{\chi'}&& \cr 
&&&&\cr &&&&\cr 
&L_\tcp^\ttp\ 
\nwarrow\searrow
\tilde L_\tcp^\ttp 
&& \tilde L_{\chi'}^\ttp\ 
\swarrow\nearrow 
L_{\chi'}^\ttp & \cr
&&&&\cr &&&&\cr 
&&\hc_\tcp^\ttp\ \simeq\ \hc_{\chi'}^\ttp&&\cr }
\ee 

\section{Positive energy UIRs of
D=6 conformal supersymmetry} 

In this section we review \cite{Dosu6}. 
The superconformal algebras in $D=6$ are ~$\cg ~=~ osp(8^*/2N)$ 
(real forms of ~$osp(8/2N) ~\cong~ D(4,N)$, \cite{Kab}). 
We label their physically relevant representations by the 
signature:
\eqn{sgn} \chi ~=~ [\, d\,;\, n_1\,,\, n_2\,,\, n_3\,; 
\,a_1\,,...,a_N\,] \ee 
where ~$d$~ is the conformal weight, ~$n_1,n_2,n_3$~ are 
non-negative integers which are Dynkin labels of the finite-dimensional 
irreps of the $D=6$ Lorentz algebra ~$so(5,1)$, and ~$a_1,...,a_N$~ are 
non-negative integers which are Dynkin labels of the finite-dimensional 
irreps of the internal (or $R$) symmetry algebra ~$usp(2N)$.
The even subalgebra of ~$osp(8^*/2N)$~ is the algebra 
~$so^*(8) \oplus usp(2N)$, and ~$so^*(8)\cong so(6,2)$~ is the 
$D=6$ conformal algebra. 

In \cite{Dosu6} we gave a constructive proof for the UIRs of ~$osp(8^*/2N)$~ 
following the methods used for the $D=4$ superconformal algebras 
~$su(2,2/N)$, cf. \cite{DPu,DPf,DPp}. The main tool
is an adaptation of the Shapovalov form on the Verma modules
~$V^\chi$~ over the complexification ~$\cg^\bac ~=~ osp(8/2N)$~
of ~$\cg$. The UIRs are realized as irreducible
factor-modules of the Verma modules ~$V^\chi$. (The reducibility
conditions of ~$V^\chi$~ are derived according to \cite{Kc}.) 
The main result is: 

\nt {\bf Theorem:}~~ All positive energy unitary irreducible 
representations of the 
conformal superalgebra ~$osp(8^*/2N)$~ characterized by the signature 
~$\chi$~ in (\ref{sgn}) are obtained for real ~$d$~ and are given 
in the following list: 
\bea
&&d ~\geq~ d^-_{11} ~=~ \half (3 n_1 + 2 n_2 + n_3 ) + 2r_1 +6 
\ , \quad {\rm no\ restrictions\ on}\ n_j \qquad \\
&&d ~=~ d^-_{21} ~=~ \half ( n_3 + 2 n_2 )+ 2r_1 + 4 
\ , \quad n_1 =0 \\
&&d ~=~ d^-_{31} ~=~ \half n_3 + 2r_1 + 2 
\ , \quad n_1 =n_2 = 0 \label{unt}\\
&&d ~=~ d^-_{41} ~=~ 2r_1 
\ , \quad n_1 =n_2 = n_3 = 0 \eea
where ~$d^-_{j1}$~ are the four distinguished reducibility points
of the Verma modules: 
\eqn{red} d^-_{j1} ~\doteq~ \half (3 n_1 + 2 n_2 + n_3 ) ~-~
2 \sum_{s=1}^{j-1}\ n_j ~+~ 2r_1 ~+~ 8 - 2j \ee

\nt {\it Remark:}~~
For ~$N=1,2$~ the Theorem was conjectured by Minwalla \cite{Min},
except that he conjectured unitarity also for the open interval
$(d^-_{31},d^-_{21})$ with conditions on $n_j$ as in (\ref{unt}).
We should note that this conjecture could be reproduced neither
by methods of conformal field theory \cite{FSa}, nor by the
oscillator method \cite{GSB} (cf. \cite{Min}), and thus was in
doubt. To compare with the notations of \cite{Min} one should
use the following substitutions: $n_1\ =\ h_2-h_3\,$, $n_2\ =\
h_1-h_2\,$, $n_3\ =\ h_2+h_3\,$, $r_1=k$, and $h_j$ are all
integer or all half-integer. The fact that $n_j\geq 0$ for
$j=1,2,3$ translates into: ~$h_1 \geq h_2 \geq |h_3|$, i.e., the
parameters $h_j$ are of the type often used for representations
of $so(2N)$ (though usually for $N\geq 4$). Note also that the
statement of the Theorem is arranged in \cite{Min} according to
the possible values of ~$n_i$~ first and then the possible values
of ~$d$. To compare with the notation of \cite{FSa} we use the
substitution ~$(n_1,n_2,n_3)\to (J_3,J_2,J_1)$. Some UIRs at the
four exceptional points ~$d^-_{i1}$~ were constructed in
\cite{GNW} by the oscillator method (some of these were
identified with Cartan-type signatures like (\ref{sgn}) in, e.g.,
\cite{Min}, \cite{EFS}).

\vskip 5mm

\nt
{\bf Acknowledgement.} ~The author would like to thank the
organizers and especially B. Dragovic for the invitation to give
lectures at the School and for the warm hospitality at Sokobanja.

\end{document}